\documentclass[12pt]{article}
\usepackage{amsfonts,euscript}
\begin{document}
\title{ Lorentz invariant supersymmetric mechanism for non(anti)commutative deformations of space-time geometry}
\author {A.A.~Zheltukhin${}^{a,b}$\\
{\normalsize ${}^a$ Kharkov Institute of Physics and Technology, 61108 Kharkov, Ukraine}\\
{\normalsize ${}^{b}$ Institute of Theoretical Physics, University of Stockholm}\\
{\normalsize  SE-10691, AlbaNova, Stockholm, Sweden}}                                            
\date{}

\maketitle

\begin{abstract}
  A supersymmetric Lorentz invariant mechanism for superspace deformations is proposed. It is based on an  extension of superspace by one $\lambda_{a}$ or several Majorana spinors  associated with the Penrose twistor picture. Some examples of Lorentz invariant
 supersymmetric Poisson and Moyal brackets are constructed and the correspondence:  $\theta_{mn}\leftrightarrow i\psi_{m}\psi_{n},\quad C_{ab}\leftrightarrow \lambda_{a}\lambda_{b},\quad \Psi^{a}_{m}\leftrightarrow \psi_{m}\lambda^{a}$ mapping the brackets depending on the constant background
 into the Lorentz covariant supersymmetric brackets is established. 
The correspondence reveals the role of the composite
anticommuting  vector $\psi_{m}=-{1\over 2}(\bar\theta\gamma_{m}\lambda)$
as a covariant measure of space-time coordinate noncommutativity.

\end{abstract}

\section{Introduction}
 
The unification of physics and  mathematics in the  development of noncommutative quantum  geometry \cite{Sny, Casal, BrSch, SchNwh, Con, Man, Kon} and field theories  \cite{BaFiShSu, CoDoSch, SeWi, FeL, 
MaSSWes, KLMa}  resulted in new  ideas and approaches (see reviews \cite{DoN}, \cite{Sza} and additional references therein). One of them has come from string theory, where the noncommutativity of the  bosonic string coordinates $x_{m}$ in the presence of the constant antisymmetric field $B_{mn}$ was observed  \cite{SeWi}.  More recently, the  noncommutativity  between  the components of the odd spinor coordinate $\theta_{a}$ in the  presence of a constant graviphoton field  $C_{ab}$ was considered  in \cite{OoVa}.  The constant  gravitino background  $\Psi^{a}_{m}$ resulted in the  noncommutativity between  the $x_{m}$ and $\theta_{a}$ coordinates \cite{KPT, BoGNwh}. These results have focused attention on  the role of constant  background fields
in superspace deformations. Studying  field/string theories and supersymmetry preservation in  the superspaces deformed by the graviphoton background  \cite{KPT}, \cite{ BoGNwh} was further advanced in \cite{Sei} and \cite{SeiBer}. A general approach to the  construction of superspace deformations in a  constant background  based on the Moyal-Weyl quantization of the Poisson  brackets  was developed in            \cite{FeL, KPT, FeLMa}. The presence of constant background fields in the much discussed deformed (anti)commutation relations for the 
(super)coordinate operators leads to the well-recognized problem of Lorentz symmetry breaking. The idea to overcome this problem by using twisted Hopf algebra was recently proposed in \cite{ChKNT} and its supersymmetric generalization was realized in \cite{KoSa} and  further developed in \cite{BZ}, \cite{IhM}. 
Another way was observed in \cite{UZnc}, where the Hamiltonian structure of free twistor-like model  \cite{ZUm} of super p-brane in $N=1$ superspace extended by tensor central charge coordinates was studied  and  the Dirac bracket-non(anti)commutativity of the brane (super)coordinates was established. The r.h.s. of these D.B's. have been  constructed from the components of auxiliary twistor-like dynamical variables which are  Lorentz covariant and supersymmetric. 
It gives  a hint that a hidden spinor structure, associated with the Penrose twistor  picture
\cite{Penrose, Ferb,Witt, Shir, BC}  might be an alternative source for the non(anti)commutativity of the quantum  space-time (super)coordinates.
Accepting such a possibility we start here from  the above mentioned spinor extension of the  $N=1\, D=4$ superspace $(x_{m},\theta_{a})$ by one commuting Majorana spinor coordinate $\lambda_{a}$ and construct Lorentz invariant and supersymmetric Poisson and Moyal brackets generating non(anti)commutative relations of the (super)coordinates. An interesting feature of these brackets is the presence of a real (or complex) Grassmannian vector $\psi_{m}$, which is well known from the theory of spinning strings and particles \cite{GSW}, in the r.h.s. of the brackets of $x_m$ with $x_n$ and $\theta_a$.
The odd vector  $\psi_{m}$ appears there in the  form 
 of an effective variable $\psi_{m}=-{1\over 2}(\bar\theta\gamma_{m}\lambda)$ \cite{VZ} composed  from the two Majorana spinors $\lambda_{a},\theta_{a}$ and encoding primordial degrees of freedom presented  by $\theta_{a}$. 
In the simplest case there is a correspondence between the Lorentz invariant brackets in question  and the known brackets including the constant  background fields. That 
 correspondence may be schematically illustrated as the map transforming the field dependent brackets into the new  brackets and vice versa :  $$B_{mn}\leftrightarrow i\psi_{m}\psi_{n},\quad C_{ab}\leftrightarrow \lambda_{a}\lambda_{b},\quad \Psi^{a}_{m}\leftrightarrow \psi_{m}\lambda^{a}.$$ (modulo the change $ B_{mn}\leftrightarrow \theta_{mn}\equiv B^{-1}_{mn}$ etc).
The schematical correspondence is preserved in the more sophisticated cases considered below and 
points to a deep correlation between  the spin structure, non(anti)commutativity and supergravity. 
We find also Lorentz invariant and supersymmetric brackets, where nonanticommutativity occurs only for the components of $\theta^{a}$ with opposite chirality. 
  The generalizations to the higher  dimensions $D=2,3,4 (mod8), N>1$ and several additional  spinors are discussed.

\section{Supersymmetry algebra in the presence of a spinor coordinate}

 Using the agreements of \cite{UZnc} we accept here the $D=4\, N=1$ supersymmetry transformation
 law in the  presence of the twistor-like Majorana spinor coordinates $(\nu_\alpha, \bar\nu_{\dot\alpha})$ in the form
\begin{equation}\label{1/8}
\begin{array}{c}
\delta\theta_\alpha=\varepsilon_\alpha,\quad 
\delta x_{\alpha\dot\alpha}=
2i(\varepsilon_{\alpha}\bar\theta_{\dot\alpha}-
\theta_{\alpha}\bar\varepsilon_{\dot\alpha}), \quad \delta\nu_\alpha=0,          
\end{array}
\end{equation}
The supercharges $Q_\alpha$ and $\bar Q_{\dot\alpha}$ of  the superalgebra (\ref{1/8}) are 
given by the differential operators 
\begin{equation}\label{2/29}
\begin{array}{c}
Q^{\alpha}=\frac{\partial}{\partial\theta_\alpha}+2i\bar\theta_{\dot\alpha}
\partial^{\alpha\dot\alpha}, \quad
{\bar Q}^{\dot\alpha}\equiv -(Q^{\alpha})^{*}=\frac{\partial}{\partial\bar\theta_{\dot\alpha}} + 2i\theta_{\alpha}\partial^{\alpha\dot\alpha},\quad
[ Q^{\alpha}, \bar Q^{\dot\alpha} ]_{+}= 4i\partial^{\alpha\dot\alpha}, 
\end{array}
\end{equation}
where $\partial^{\alpha\dot\alpha} \equiv 
\frac{\partial}{\partial x_{\alpha\dot\alpha}}$  and the correspondent supersymmetric covariant derivatives $D,\bar D$ are 
\begin{equation}\label{3/30}
\begin{array}{c}
D^{\alpha}=\frac{\partial}{\partial\theta_\alpha}-2i\bar\theta_{\dot\alpha}
\partial^{\alpha\dot\alpha}, \quad
{\bar D}^{\dot\alpha}\equiv -(D^{\alpha})^{*}=\frac{\partial}{\partial\bar\theta_{\dot\alpha}}- 2i\theta_{\alpha}\partial^{\alpha\dot\alpha},
 \quad  
[ D^{\alpha},\bar D^{\dot\beta}]=-4i\partial^{\alpha\dot\alpha},
\\[0.2cm]
[ Q^{\alpha}, D^{\beta}]_{+}= [ Q^{\alpha}, \bar D^{\dot\beta}]_{+}=
{[ \bar Q}^{\dot\alpha}, D^{\beta}]_{+}={ [\bar Q}^{\dot\alpha},\bar D^{\dot\beta}]_{+}=0.
\end{array}
\end{equation}
The spinor coordinates $ (\nu_{\alpha}, \bar\nu_{\dot\alpha})$ and the light-like real vector $\varphi_{\alpha\dot\alpha}$ composed from them  
\begin{equation}\label{4/4}
\varphi_{\alpha\dot\alpha}\equiv  \nu_{\alpha}\bar\nu_{\dot\alpha}, \quad
\varphi_{\alpha\dot\alpha}\nu^{\alpha}=
\varphi_{\alpha\dot\alpha}\bar\nu^{\dot\alpha}=0, 
\quad \delta\varphi_{\alpha\dot\beta}=0\quad
\end{equation}
may be  used  to construct the  Lorentz invariant differential operators 
$ D, \bar D, \partial$
\begin{equation}\label{5/32}
\begin{array}{c}
D= \nu_{\alpha}D^{\alpha},\quad \bar D= \bar\nu_{\dot\alpha}\bar {D^{\dot\alpha}},\quad  \partial=\varphi_{\alpha\dot\alpha} \partial^{\alpha\dot\alpha}
\end{array}
\end{equation}
which form a supersymmetric subalgebra of the algebra of derivatives
\begin{equation}\label{6}
\begin{array}{c} 
[D, \bar D]_{+}=-4i\partial,\quad [D, D]_{+}=[\bar D, \bar D]_{+}=0, \quad [D,\partial]=
 [\bar D,\partial]=[\partial,\partial]=0.
\end{array}
\end{equation}
Introducing $D_{\pm}$ combinations of the invariant derivatives $D, \bar D$ 
\begin{equation}\label{7}
\begin{array}{c} 
D_{\pm}\equiv D\pm \bar D
\end{array}
\end{equation}
one can split the Lorentz invariant complex subalgebra (\ref{6}) into two invariant and (anti)commuting subalgebras formed by the generators   $(D_{-},\partial)$ and $(D_{+},\partial) $
\begin{equation}\label{8/36}
\begin{array}{c} 
[D_{\pm}, D_{\pm}]_{+}=\mp8i\partial,\quad [D_{+}, D_{-}]_{+}=0, \quad 
[ D_{\pm}, \partial]=[\partial,\partial]=0.
\end{array}
\end{equation}

A twistor-like character of the Majorana spinor  $(\nu_\alpha, \bar\nu_{\dot\alpha})$ means that its dilatations, generated by the differential operator $\Delta$
\begin{equation}\label{9/36}
\begin{array}{c}
\Delta=\nu_{\alpha}
\frac{\partial}{\partial\nu_\alpha} + \bar\nu_{\dot\alpha}\frac {\partial}{\partial\bar\nu_{\dot\alpha}},
\end{array}
\end{equation}
have to be an additional symmetry of  physical theories of massless fields. Taking into account this dilaton symmetry assumes  extension of the superalgebra (\ref{6}) by the  dilaton generator $\Delta$
\begin{equation}\label{10/36}
\begin{array}{c} 
[D, \bar D]_{+}=-4i\partial,\quad [D, D]_{+}=[\bar D, \bar D]_{+}=0,
\\[0.2cm]
[\Delta,D]= D,\quad[\Delta,\bar D]=\bar D,\quad[\Delta,\partial]=2\partial,\\[0.2cm] 
[\partial, D]=[\partial,\bar D]=[\partial,\partial]=[\Delta,\Delta]=0,
\end{array}
\end{equation}
which has two real anticommutative  subalgebras formed by the generators $(D_{\pm},\partial,\Delta )$ \begin{equation}\label{11/37}
\begin{array}{c} 
[D_{\pm}, D_{\pm}]_{+}=\mp8i\partial,
\quad 
[\Delta,D_{\pm}]= D_{\pm},\quad 
[\Delta,\partial]=2\partial,
\\[0.2cm] 
[D_{+}, D_{-}]_{+}=[ D_{\pm},\partial]=[\partial,\partial]=[\Delta,\Delta]=0.
\end{array}
\end{equation} 

The Lorentz invariant supersymmetric differential operators forming the superalgebras (\ref{10/36}), (\ref{11/37}) may be used as building blocks for the construction of  Lorentz invariant and supersymmetric non(anti)commutative relations among quantum operators of (super)coordinates.

\section{Lorentz invariant supersymmetric Poisson brackets: non(anti)commutativity of space-time coordinates}

To clarify  the  role of $(\nu_\alpha, \bar\nu_{\dot\alpha})$ in the formation of Lorentz invariant non(anti)commutative relations among $x_{\alpha\dot\alpha},\theta_{\alpha}, \bar\theta_{\dot\alpha}$ we consider  the Poisson bracket  constructed from the three differential operators $(D_{-},\partial,\Delta )$  forming the simplest superalgebra (\ref{11/37})
\begin{equation}\label{13/31}
\begin{array}{c} 
\{ F, G \}= F\,[\, -\frac{i}{4}\,
{\stackrel{\leftarrow}{D}}_{-}
{\stackrel{\rightarrow}{D}}_{-}+ (\stackrel{\leftarrow}{\partial}\stackrel{\rightarrow}
{\Delta} - \stackrel{\leftarrow}{\Delta}\stackrel{\rightarrow}{\partial})\,]\,G,
\end{array}
\end{equation}
where $ \{  , \}_{P.B.}\equiv \{  , \}$ and $F(x,\theta,\bar\theta,\nu,\bar\nu), G(x,\theta,\bar\theta,\nu,\bar\nu) $ are  generalized superfields depending on both the superspace coordinates $( x,\theta,\bar\theta)$ and on the spinor coordinates $(\nu, \bar\nu)$.
The Lorentz invariant and supersymmetric differential operators  ${\stackrel{\leftarrow}{D}}_{-}, \stackrel{\leftarrow}{\partial},\stackrel{\leftarrow}{\Delta}$ define derivatives acting from the right hand side.
 Conversely, the differential operators (\ref{11/37}) act  from the l.h.s  and 
coincide with the left  derivatives ${\stackrel{\rightarrow}{D}}_{-},\stackrel{\rightarrow}{\partial},\stackrel{\rightarrow}{\Delta}$
\begin{equation}\label{14}
\begin{array}{c} 
{\stackrel{\rightarrow}{D}}_{-}G \equiv D_{-}G, \quad 
\stackrel{\rightarrow}{\partial}G\equiv\partial G, \quad
\stackrel{\rightarrow}{\Delta}G \equiv\Delta G
\end{array}
\end{equation}
The left and right invariant derivatives  in (\ref{13/31}) are  connected by the relations 
\begin{equation}\label{15/35}
\begin{array}{c} 
F\stackrel{\leftarrow}{D} =(-1)^{f}\stackrel{\rightarrow}{D}F, 
\quad
F\stackrel{\leftarrow}{\bar D}=(-1)^{f}
\stackrel{\rightarrow}{\bar D}F,
\quad
F\stackrel{\leftarrow}{\partial}=\stackrel{\rightarrow}{\partial}F, 
\quad
 F\stackrel{\leftarrow}{\Delta}=\stackrel{\rightarrow}{\Delta}F,
\end{array}
\end{equation}
where  ${f}=0,1$ is the  Grassmannian grading of the superfield $F$.

The action of $D, \bar D,  D_{\pm}$  on the composite coordinates 
$\psi$ and $\varphi $ is given by the relations 
\begin{equation}\label{16/33}
\begin{array}{c} 
\stackrel{\rightarrow}{D}\psi_{\alpha\dot\alpha}=-i\varphi_{\alpha\dot\alpha}, \quad 
\stackrel{\rightarrow}{\bar D}\psi_{\alpha\dot\alpha}=i\varphi_{\alpha\dot\alpha}, 
\quad 
\psi_{\alpha\dot\alpha}\equiv i(\nu_{\alpha}\bar\theta_{\dot\alpha}- \theta_{\alpha}\bar\nu_{\dot\alpha}),
\\[0.2cm]
\stackrel{\rightarrow}{D_{-}}\psi_{\alpha\dot\alpha}=-2i\varphi_{\alpha\dot\alpha},\quad 
\stackrel{\rightarrow}{D_{+}}\psi_{\alpha\dot\alpha}=\stackrel{\rightarrow}{D_{\pm}}\varphi_{\alpha\dot\alpha}=0.
\end{array}
\end{equation}
After the substitions of the (super)coordinates under discussion in the P.B. (\ref{13/31}) we find  non(anti)commutative P.B's. for  them. 

The twistor-like coordinates have zero P.B's. among themselves  
\begin{equation}\label{17/1}
\{\nu_{\alpha}, \nu_\beta\}=\{\nu_\alpha,\bar\nu_{\dot\beta}\}=
\{\bar\nu_\alpha, \bar\nu_{\dot\beta}\}=0
\end{equation}
and  with the Grassmannian spinors $(\theta_\alpha, \bar\theta_{\dot\alpha})$ 
\begin{equation}\label{18/2}
\begin{array}{c}
\{\nu_\alpha,\theta_\beta\}=
\{\nu_\alpha, \bar\theta_{\dot\beta}\}=
\{ \bar\nu_{\dot\alpha},\theta_\beta \}=
\{ \bar\nu_{\dot\alpha},\bar\theta_{\dot\beta}\}=0.
\end{array}
\end{equation}
However, they have  non zero P.B's.  with  the space-time coordinates  $x_{\alpha\dot\alpha}$ 
\begin{equation}\label{19/3}
\{ x_{\alpha\dot\alpha}, \nu_\beta \}=\varphi_{\alpha\dot\alpha}\nu_\beta,
\quad
\{ x_{\alpha\dot\alpha}, \bar\nu_{\dot\beta} \}=\varphi_{\alpha\dot\alpha}\bar\nu_{\dot\beta},
\end{equation}
The remaining non zero P.B's. define the  P.B's. among  the space-time coordinates  $x_{\alpha\dot\alpha}$  and spinors $(\theta_\alpha, \bar\theta_{\dot\alpha})$
\begin{equation}\label{20/7}
\begin{array}{c}
\{x_{\alpha\dot\alpha},x_{\beta\dot\beta}\}=-i\psi_{\alpha\dot\alpha}
\psi_{\beta\dot\beta},
\\[0.2cm]
\{x_{\alpha\dot\alpha},\theta_\beta \}= \frac{i}{2}
\psi_{\alpha\dot\alpha}\nu_{\beta},
\quad
\{x_{\alpha\dot\alpha},\bar\theta_{\dot\beta} \}=-\frac{i}{2}
\psi_{\alpha\dot\alpha}\bar\nu_{\dot\beta},
\\[0.2cm]
\{\theta_{\alpha},\theta_{\beta} \}=\frac{i}{4}\varphi_{\alpha\beta}, 
\quad
\{\theta_\alpha,\bar\theta_{\dot\beta} \}=-\frac{i}{4}\varphi_{\alpha\dot\beta}, 
\quad
\{\bar\theta_{\dot\alpha}, \bar\theta_{\dot\beta}\}=\frac{i}{4}\bar\varphi_{\dot\alpha\dot\beta},
\end{array}
\end{equation}
where  $\varphi_{\alpha\beta}, \bar\varphi_{\dot\alpha\dot\beta}$ are composite symmetric spin-tensors 
\begin{equation}\label{21/6}
\varphi_{\alpha\beta}\equiv\nu_{\alpha}\nu_{\beta}, \quad
\bar\varphi_{\dot\alpha\dot\beta}\equiv  \bar\nu_{\dot\alpha}\bar\nu_{\dot\beta},
\quad \delta\varphi_{\alpha\beta}=\delta\bar\varphi_{\dot\alpha\dot\beta}=0
\end{equation} 
orthogonal to  the vector $\varphi_{\alpha\dot\alpha}$ (\ref{4/4}) and to the composite Grassmannian vector $\psi_{\alpha\dot\alpha}$
\begin{equation}\label{22/5}
\psi_{\alpha\dot\alpha}\equiv i(\nu_{\alpha}\bar\theta_{\dot\alpha}- 
\theta_{\alpha}\bar\nu_{\dot\alpha}),  
\quad 
\varphi^{\alpha\dot\alpha}\psi_{\alpha\dot\alpha}= \varphi^{\alpha\beta}\psi_{\alpha\dot\alpha}= \bar\varphi^{\dot\alpha\dot\beta}\psi_{\alpha\dot\alpha}=0,
\quad 
\delta\psi_{\alpha\dot\alpha}=-i(\varepsilon_\alpha \bar\nu_{\dot\alpha}-
\bar\varepsilon_{\dot\alpha}\nu_\alpha)
\end{equation}

The appearance of the odd vector $\psi_{\alpha\dot\alpha}$ (\ref{22/5}) associated with  the description of the spin degrees of freedom of fermions in the r.h.s. of  P.B's. (\ref{20/7}) hints  on a spin structure of superspaces in back of the coordinate's non(anti)commutativity\footnote{Let us remind that composite character of the anticommuting vector $\psi_{\alpha\dot\alpha}$ (\ref{22/5}) was revealed in  \cite{VZ}, where the spinor representation (\ref{22/5}) was found to be the general solution of Dirac constraints $p^{\alpha\dot\alpha}\psi_{\alpha\dot\alpha}=0=p^{\alpha\dot\alpha}p_{\alpha\dot\alpha}$ characterising massless spinning particle \cite{BDZDH},\cite{Zkk}.
 This spinor representation was  important to find equivalence between spinning and Brink-Schwarz  superparticles.}.
  The Lorentz covariance of the Poisson brackets (\ref{17/1})-(\ref{20/7}) is provided by the spinor, vector and spin-tensor representations of the Lorentz group involved in the r.h.s. of the  Poisson brackets. These P.B's. are also supersymmetric by the construction. 

In the next Section we prove the Jacobi identities for the P.B's. (\ref{17/1})-(\ref{20/7}). 
 
\section{Proof of the Jacobi identities}

The graded Jacobi identities for the considered P.B. algebra have the standard form 
\begin{equation}\label{23/9}
\{\{A,B \},C\} +(-1)^{(b+c)a}\{\{B,C \},A\}
+ (-1)^{c(a+b)}\{\{C,A \},B\}=0,
\end{equation}
where $a=0,1$ is the Grassmannian  grading of $A$.
To  prove  these identities for the P.B's. (\ref{17/1})-(\ref{20/7}) one needs to study the  Poisson brackets of the composite vectors $ \varphi_{\alpha\dot\beta}$  (\ref{4/4}),   $\psi_{\alpha\dot\alpha}$ (\ref{22/5}) and spin-tensors  $ \varphi_{\alpha\beta}, \bar\varphi_{\dot\alpha\dot\beta}$  (\ref{21/6}) between themselves and with $ x_{\alpha\dot\alpha},\theta_{\alpha},  \bar\theta_{\dot\alpha}$. The P.B's. (\ref{17/1}), (\ref{18/2}) together with the definitions (\ref{21/6}), (\ref{22/5}) show the P.B.-commutativity of $ \varphi_{\alpha\beta}, \varphi_{\alpha\dot\beta}, \bar\varphi_{\dot\alpha\dot\beta}$
 between themselves and with  $(\nu_{\alpha}, \bar\nu_{\dot\alpha})$, $(\theta_{\alpha}, \bar\theta_{\dot\alpha}) $  and $\psi_{\alpha\dot\alpha}$
\begin{equation}\label{24/10}
\begin{array}{c}
\{\varphi_{\ast\ast},\nu_{\alpha}\}=\{\varphi_{\ast\ast}, \bar\nu_{\dot\alpha}\}=\{\psi_{\alpha\dot\alpha},\nu_{\beta}\}=
 \{\psi_{\alpha\dot\alpha},\bar\nu_{\dot\beta}\}=0,
\\[0.2cm]
\{\varphi_{\ast\ast},\varphi_{\ast\ast}\}=
\{\varphi_{\ast\ast},\theta_{\alpha}\}=
\{\varphi_{\ast\ast},\bar\theta_{\dot\alpha}\}=
\{\varphi_{\ast\ast},\psi_{\gamma\dot\gamma}\}=0,
\end{array}
\end{equation}
where $\varphi_{\ast\ast}\equiv ( \varphi_{\alpha\beta}, \varphi_{\alpha\dot\beta}, \bar\varphi_{\dot\alpha\dot\beta})$ is a condenced symbol for the
 composite  coordinates (\ref{4/4}) and (\ref{21/6}). However, the P.B. of the spin-tensors $\varphi_{\alpha\beta}, \varphi_{\alpha\dot\beta}, \bar\varphi_{\dot\alpha\dot\beta}$ with  $x_{\gamma\dot\gamma}$ are different from zero
\begin{equation}\label{25/11}
\{x_{\alpha\dot\alpha},\varphi_{\beta\gamma}\}=2\varphi_{\alpha\dot\alpha}
\varphi_{\beta\gamma},
\quad 
\{x_{\alpha\dot\alpha},\varphi_{\beta\dot\gamma}\}=2\varphi_{\alpha\dot\alpha}\varphi_{\beta\dot\gamma},
\quad 
\{x_{\alpha\dot\alpha},\bar\varphi_{\dot\beta\dot\gamma}\}=2\varphi_{\alpha\dot\alpha}\bar\varphi_{\dot\beta\dot\gamma},
\end{equation}
as well as, the P.B. between $ x_{\alpha\dot\alpha},\psi_{\beta\dot\beta}$ and $(\theta_{\beta},\bar\theta_{\dot\beta}) $  
\begin{equation}\label{26/12}
\begin{array}{c}
\{x_{\alpha\dot\alpha},\psi_{\beta\dot\beta}\}=\varphi_{\alpha\dot\alpha}
\psi_{\beta\dot\beta}+ \varphi_{\beta\dot\beta}\psi_{\alpha\dot\alpha},
\\[0.2cm]
\{\psi_{\alpha\dot\alpha},\psi_{\beta\dot\beta}\}=
-i\varphi_{\alpha\dot\alpha}\varphi_{\beta\dot\beta},
\\[0.2cm]
\{\psi_{\alpha\dot\alpha}, \theta_{\beta}\}=\frac{1}{2}
\varphi_{\alpha\dot\alpha}\nu_{\beta}, \quad
\{\psi_{\alpha\dot\alpha}, \bar\theta_{\dot\beta}\}=-\frac{1}{2}
\varphi_{\alpha\dot\alpha}\bar\nu_{\dot\beta}.
\end{array}
\end{equation}
A combination of the  P.B. relations (\ref{24/10}) together with ones (\ref{26/12}) results in the relation  
\begin{equation}\label{27/13}
\{\{\psi_{\alpha\dot\alpha},\psi_{\beta\dot\beta}\},\psi_{\gamma\dot\gamma}\}=0
\end{equation}
which proves the graded Jacobi identity (\ref{13/31}) for the case $A=B=C=\psi$
\begin{equation}\label{28/14}
Cycle\{\{\psi_{\alpha\dot\alpha},\psi_{\beta\dot\beta}\},\psi_{\gamma\dot\gamma}\}=0
\end{equation}
The same result occurs for the Jacoby cycles cubic in $(\theta_{\alpha},
\bar\theta_{\dot\alpha})$
\begin{equation}\label{29/15}
Cycle\{\{\theta_{\alpha},\theta_{\beta}\},\theta_{\gamma}\}=
...= Cycle\{\{ \bar\theta_{\dot\alpha},\bar\theta_{\dot\beta}\},
\bar\theta_{\dot\gamma}\}=0,
\end{equation}
as well as, for the cycles  quadratic in $\theta_{\alpha}$ or $\psi_{\alpha\dot\alpha}$ and linear in 
$(\nu_{\gamma},\bar\nu_{\dot\gamma})$ 
\begin{equation}\label{30/16}
\begin{array}{c}
Cycle\{\{\theta_{\alpha},\theta_{\beta}\},\nu_{\gamma}\}=....=
Cycle\{\{\bar\theta_{\dot\alpha},\bar\theta_{\dot\beta}\},
\bar\nu_{\dot\gamma}\}=0,
\\[0.2cm]
Cycle\{\{\psi_{\alpha\dot\alpha},\psi_{\beta\dot\beta}\},\nu_{\gamma}\}=
Cycle\{\{\psi_{\alpha\dot\alpha},\psi_{\beta\dot\beta}\},\bar\nu_{\dot\gamma}\}
=0
\end{array}
\end{equation}
and for other trivial Jacobi cycles cubic or quadratic in $ (\nu_{\gamma},\bar\nu_{\dot\gamma})$ and linear in $ (\theta_{\alpha},\bar\theta_{\dot\alpha})$ or $\psi_{\alpha\dot\alpha}$. 

To calculate the Jacobi cycle cubic in $x_{\alpha\dot\alpha}$ we use the relation
\begin{equation}\label{31/17}
\{\{x_{\alpha\dot\alpha},x_{\beta\dot\beta}\},x_{\gamma\dot\gamma}\}=2i(\psi_{\alpha\dot\alpha}\psi_{\beta\dot\beta})\varphi_{\gamma\dot\gamma}+ i(\psi_{\alpha\dot\alpha}\varphi_{\beta\dot\beta}-\psi_{\beta\dot\beta}\varphi_{\alpha\dot\alpha})\psi_{\gamma\dot\gamma}
\end{equation}
arisen from the P.B's. (\ref{20/7}) and (\ref{26/12}) and resulting in  zero Jacobi cycle  
\begin{equation}\label{32/18}
Cycle\{\{x_{\alpha\dot\alpha},x_{\beta\dot\beta}\},x_{\gamma\dot\gamma}\}=0.
\end{equation}
It follows from the mutual cancellation between the contributions of first and last summands in the r.h.s. of the cyclic sum generated by Eq. (\ref{31/17}). Next one can see that the Jacobi cycles quadratic in $x_{\alpha\dot\alpha}$ and linear in $\psi_{\alpha\dot\alpha}$ or $(\theta_{\alpha},\bar\theta_{\dot\alpha})$  are  equal to zero 
\begin{equation}\label{33/19}
\begin{array}{c}
Cycle\{\{x_{\alpha\dot\alpha},x_{\beta\dot\beta}\},\psi_{\gamma\dot\gamma}\}
=0,
\\[0.2cm]
Cycle\{\{x_{\alpha\dot\alpha},x_{\beta\dot\beta}\},\theta_{\gamma}\}
=Cycle\{\{x_{\alpha\dot\alpha},x_{\beta\dot\beta}\},\bar\theta_{\dot\gamma}\}= 0,
\end{array}
\end{equation}
because of  the relations 
\begin{equation}\label{34/20}
\begin{array}{c}
\{\{x_{\alpha\dot\alpha},x_{\beta\dot\beta}\},
\psi_{\gamma\dot\gamma}\}
=-(\psi_{\alpha\dot\alpha}\varphi_{\beta\dot\beta}- 
\varphi_{\alpha\dot\alpha}\psi_{\beta\dot\beta})\varphi_{\gamma\dot\gamma},
\\[0.2cm]
\{\{x_{\beta\dot\beta},\psi_{\gamma\dot\gamma}\},
x_{\alpha\dot\alpha}\} - (\alpha\leftrightarrow\beta)=
(\psi_{\alpha\dot\alpha}\varphi_{\beta\dot\beta}- 
\varphi_{\alpha\dot\alpha}\psi_{\beta\dot\beta})\varphi_{\gamma\dot\gamma},
\\[0.2cm]
\{\{x_{\alpha\dot\alpha},x_{\beta\dot\beta}\},\theta_{\gamma}\}=
-\frac{i}{2}(\psi_{\alpha\dot\alpha}\varphi_{\beta\dot\beta}-
\varphi_{\alpha\dot\alpha}\psi_{\beta\dot\beta})\nu_{\gamma},
\\[0.2cm]
\{\{ x_{\beta\dot\beta},\theta_{\gamma}\}, x_{\alpha\dot\alpha}\} - (\alpha\leftrightarrow\beta)= 
\frac{i}{2}(\psi_{\alpha\dot\alpha}\varphi_{\beta\dot\beta}- 
\varphi_{\alpha\dot\alpha}\psi_{\beta\dot\beta})\nu_{\gamma}
\end{array}
\end{equation}
and their complex conjugate following from the P.B's. (\ref{19/3}), (\ref{20/7}) and (\ref{26/12}).
A similar cancellation takes place in  the Jacobi cycles quadratic in $\psi_{\alpha\dot\alpha}$ and linear in  $x_{\alpha\dot\alpha}$ or $\theta_{\alpha},\bar\theta_{\dot\alpha}$
\begin{equation}\label{35/21}
\begin{array}{c}
Cycle\{\{\psi_{\alpha\dot\alpha},\psi_{\beta\dot\beta}\},x_{\gamma\dot\gamma}\}=0,
\\[0.2cm]
Cycle\{\{\psi_{\alpha\dot\alpha},\psi_{\beta\dot\beta}\},\theta_{\gamma}\}
=Cycle\{\{\psi_{\alpha\dot\alpha},\psi_{\beta\dot\beta}\},\bar\theta_{\dot\gamma}\}= 0,
\end{array}
\end{equation}
as it follows from the  P.B. relations 
\begin{equation}\label{36/22}
\begin{array}{c}
\{\{\psi_{\alpha\dot\alpha},\psi_{\beta\dot\beta}\},
x_{\gamma\dot\gamma}\}=4i\varphi_{\alpha\dot\alpha}\varphi_{\beta\dot\beta}
\varphi_{\gamma\dot\gamma},
\\[0.2cm]
\{\{x_{\gamma\dot\gamma},\psi_{\beta\dot\beta}\},
\psi_{\alpha\dot\alpha}\} + (\alpha\leftrightarrow\beta)=
4i\varphi_{\alpha\dot\alpha}\varphi_{\beta\dot\beta}
\varphi_{\gamma\dot\gamma},
\\[0.2cm]
\{\{\psi_{\alpha\dot\alpha},\psi_{\beta\dot\beta}\},\theta_{\gamma}\}
= \{\{\psi_{\alpha\dot\alpha},\theta_{\gamma}\},\psi_{\beta\dot\beta}\}=0.
\end{array}
\end{equation}
 Next we prove the Jacobi identites for cycles quadratic in  $\theta_{\alpha},\bar\theta_{\dot\alpha}$ and linear in $ x_{\alpha\dot\alpha} $ or   $\psi_{\alpha\dot\alpha} $ 
\begin{equation}\label{37/23}
\begin{array}{c}
Cycle\{\{\theta_{\alpha},\theta_{\beta}\}, x_{\gamma\dot\gamma}\}=\{\{\theta_{\alpha},\theta_{\beta}\}, \psi_{\gamma\dot\gamma}\}=0,
\\[0.2cm]
Cycle\{\{\theta_{\alpha},\bar \theta_{\dot\beta}\}, x_{\gamma\dot\gamma}\}=\{\{\theta_{\alpha},\bar \theta_{\dot\beta}\}, \psi_{\gamma\dot\gamma}\}=0
\end{array}
\end{equation}
and for their complex  conjugate using the relations 
\begin{equation}\label{38/24}
\begin{array}{c}
\{\{\theta_{\alpha},\theta_{\beta}\}, x_{\gamma\dot\gamma}\}=-\frac{i}{2}
\varphi_{\alpha\beta}\varphi_{\gamma\dot\gamma}, \quad
\{\{\theta_{\alpha},\bar\theta_{\dot\beta}\}, x_{\gamma\dot\gamma}\}=\frac{i}{2}\varphi_{\alpha\dot\beta}\varphi_{\gamma\dot\gamma},
\\[0.2cm]
\{\{ x_{\gamma\dot\gamma},\theta_{\beta}\}, \theta_{\alpha}\}+(\alpha\leftrightarrow\beta)= \frac{i}{2}\varphi_{\alpha\beta}\varphi_{\gamma\dot\gamma},
\\[0.2cm]
\{\{ x_{\gamma\dot\gamma},\bar\theta_{\dot\beta}\}, \theta_{\alpha}\} + \{\{ x_{\gamma\dot\gamma}, \theta_{\alpha}\},\bar\theta_{\dot\beta}\}
= \frac{i}{2}\varphi_{\alpha\dot\beta}\varphi_{\gamma\dot\gamma}.
\end{array}
\end{equation}
together with the  relations 
\begin{equation}\label{39/25}
\begin{array}{c}
\{\{\theta_{\alpha},\theta_{\beta}\}, \psi_{\gamma\dot\gamma}\}=
\{\{\theta_{\alpha}, \psi_{\gamma\dot\gamma}\},\theta_{\beta}\}=0,
\\[0.2cm]
\{\{\theta_{\alpha},\bar\theta_{\dot\beta}\}, \psi_{\gamma\dot\gamma}\}=
\{\{\theta_{\alpha}, \psi_{\gamma\dot\gamma}\},\bar\theta_{\dot\beta}\}=0
\end{array}
\end{equation}
and their complex  conjugate. The remaining nontrivail  and also  vanishing Jacobi cycles are  formed by any three coordinates from  the set  $ [x_{\alpha\dot\alpha},\psi_{\alpha\dot\alpha},(\theta_{\alpha},\bar\theta_{\dot\alpha}),(\nu_{\alpha},\bar\nu_{\dot\alpha})]$
\begin{equation}\label{40/26}
\begin{array}{c}
Cycle\{\{ x_{\alpha\dot\alpha}, \psi_{\beta\dot\beta}\},\theta_{\gamma}\}=
Cycle\{\{ x_{\alpha\dot\alpha}, \psi_{\beta\dot\beta}\},\nu_{\gamma}\}=0,
\\[0.2cm]
Cycle\{\{ x_{\alpha\dot\alpha},\theta_{\beta}\},\nu_{\gamma}\}=
Cycle\{\{ \psi_{\alpha\dot\alpha},\theta_{\beta}\},\nu_{\gamma}\}=0.
\end{array}
\end{equation}
 Their complex conjugate cycles  equal zero  too.. The proof of first and second Jacobi identities in (\ref{40/26}) is  based on the P.B. relations 
\begin{equation}\label{41/27}
\begin{array}{c}
\{\{ x_{\alpha\dot\alpha}, \psi_{\beta\dot\beta}\},\theta_{\gamma}\}=
-\frac{2}{3}\{\{ \psi_{\beta\dot\beta},\theta_{\gamma}\},
 x_{\alpha\dot\alpha}\}=2\{\{ x_{\alpha\dot\alpha},
\theta_{\gamma}\},
\psi_{\beta\dot\beta}\}
=\varphi_{\alpha\dot\alpha}\varphi_{\beta\dot\beta}\nu_{\gamma},
\\[0.2cm]
\{\{ x_{\alpha\dot\alpha}, \psi_{\beta\dot\beta}\},\nu_{\gamma}\}=
\{\{ \psi_{\beta\dot\beta},\nu_{\gamma}\}, x_{\alpha\dot\alpha}\}=
\{\{\nu_{\gamma}, x_{\alpha\dot\alpha}\}, \psi_{\beta\dot\beta}\}=0.
\end{array}
\end{equation}
The proof of third and fourth Jacobi  identitities in (\ref{40/26}) uses the P.B. relations 
\begin{equation}\label{42/28}
\begin{array}{c}
\{\{ x_{\alpha\dot\alpha},\theta_{\beta}\},\nu_{\gamma}\}=\{\{\theta_{\beta},\nu_{\gamma}\},x_{\alpha\dot\alpha}\}=\{\{\nu_{\gamma}, x_{\alpha\dot\alpha}\},\theta_{\beta}\}=0,
\\[0.2cm]
\{\{ \psi_{\alpha\dot\alpha},\theta_{\beta}\},\nu_{\gamma}\}=\{\{\theta_{\beta},\nu_{\gamma}\},\psi _{\alpha\dot\alpha}\}=\{\{\nu_{\gamma}, \psi_{\alpha\dot\alpha}\},\theta_{\beta}\}=0
\end{array}
\end{equation}
which follow from the P.B. relations (\ref{18/2}), (\ref{20/7}), (\ref{24/10}), (\ref{26/12}). 
It complets the proof of the Jacobi identities for the above introduced Lorentz  invariant Poisson brackets. The next step is to use them for the construction of the Moyal brackets.  

\section{Lorentz invariant and supersymmetric star product}

A transition to the quantum picture based on the P.B. (\ref{13/31}) may be done  by using the well known Weyl-Moyal correspondence which establishes one to one correspondence between quantum field operators and their symbols acting on commutative space-time. Then the quantum information is encoded in the  change of the usual product by the Moyal $\star$-product of their Weyl symbols.  
To realise this prescription here we  note that the 
P.B. (\ref{13/31}) may be presented as
\begin{equation}\label{43/38}
\begin{array}{c} 
\{ F, G \}=F\stackrel{\leftarrow}{{\cal D}_{\Lambda}}C^{\Lambda\Sigma}
\stackrel{\rightarrow}{{\cal D}_{\Sigma}}G=
\\[0.2cm]
F(\stackrel{\leftarrow}{{\partial}},
\stackrel{\leftarrow}{{\Delta}},
\stackrel{\leftarrow}{D_{-}})\left(
\begin{array}{lrc}0&1&0\\
-1&0&0\\0&0&\frac{-i}{4}\\
\end{array}\right)
\left(\begin{array}{cc}
\stackrel{\rightarrow}{{\partial}}&\\
\stackrel{\rightarrow}{{\Delta}}&\\
\stackrel{\rightarrow}{D_{-}}&\\\end{array}
\right)G, 
\end{array}
\end{equation}
where  the condenced notation ${\cal D}_{\Lambda}=(D_{-},\partial,\Delta )$ was used 
for the invariant derivatives of  the (-)-superalgebra (\ref{11/37}) numerated  by the  index  $\Lambda $ running  over the even and odd variables.  As a result, the superalgebra (\ref{11/37}) is presented in a condenced form 
\begin{equation}\label{44/40}
\begin{array}{c} 
[{\cal D}_{\Lambda}, {\cal D}_{\Sigma}\}= 
{\cal C}_{\Lambda\Sigma}{}^{\Xi}\,{\cal D}_{\Xi},
\end{array}
\end{equation} 
where  ${\cal C}_{\Lambda\Sigma}{}^{\Xi}$ are  the structural constants defined by the explicit  (anti)commutation  relations (\ref{11/37}) 
   and $C^{\Lambda\Sigma}$  is represented by the ${3\times3}$  matrice
\begin{equation}\label{45/39}
C^{\Lambda\Sigma}= \left(
\begin{array}{lrc}0&1&0\\
-1&0&0\\0&0&-\frac{i}{4}\\
\end{array}
\right).
\end{equation} 

The representation (\ref{43/38}) defines the  Moyal $\star$-product of the superfields $F$ and $G$ 
\begin{equation}\label{46/41}
F{\star G}= F e^{\frac{1}{2}\stackrel{\leftarrow}{{\cal D}_{\Lambda}}C^{\Lambda\Sigma}
\stackrel{\rightarrow}{{\cal D}_{\Sigma}}} G,
\end{equation}
where the  Planck constant and the velocity of light are chosen to be equal to unit. 
The  definition 
(\ref{46/41}) together with  (\ref{13/31}) yield the Moyal products  of the  (super)coordinates 
\begin{equation}\label{47/42}
\begin{array}{c} 
x_{\alpha\dot\alpha}\star x_{\beta\dot\beta}=x_{\alpha\dot\alpha}x_{\beta\dot\beta}-\frac{i}{2}\psi_{\alpha\dot\alpha}
\psi_{\beta\dot\beta},
\\[0.2cm]
x_{\alpha\dot\alpha}\star \theta_\beta = x_{\alpha\dot\alpha}\theta_{\beta} +
\frac{i}{4}
\psi_{\alpha\dot\alpha}\nu_{\beta},
\quad
x_{\alpha\dot\alpha}\star\bar\theta_{\dot\beta}=x_{\alpha\dot\alpha}\bar\theta_{\dot\beta}
-\frac{i}{4}
\psi_{\alpha\dot\alpha}\bar\nu_{\dot\beta},
\\[0.2cm]
\theta_{\alpha}\star \theta_{\beta} =\theta_{\alpha}\theta_{\beta}  +\frac{i}{8}\varphi_{\alpha\beta}, \quad
\theta_\alpha\star\bar\theta_{\dot\beta} =\theta_\alpha\bar\theta_{\dot\beta}   -\frac{i}{8}\varphi_{\alpha\dot\beta}, 
\\[0.2cm]
\bar\theta_{\dot\alpha}\star \bar\theta_{\dot\beta}=\bar\theta_{\dot\alpha}\bar\theta_{\dot\beta} +\frac{i}{8}{\bar\varphi}_{\dot\alpha\dot\beta}.
\end{array}
\end{equation} 
  Consequently, the (anti)commutators of the coordinate operators are replaced by the following  Lorentz invariant and supersymmetric Moyal brackets  
\begin{equation}\label{48/43}
\begin{array}{c} 
[x_{\alpha\dot\alpha},x_{\beta\dot\beta}]_{\star}\equiv x_{\alpha\dot\alpha}\star x_{\beta\dot\beta}- x_{\beta\dot\beta}\star x_{\alpha\dot\alpha}
=-i\psi_{\alpha\dot\alpha}
\psi_{\beta\dot\beta},
\\[0.2cm]
[x_{\alpha\dot\alpha},\theta_\beta]_{\star}= \frac{i}{2}
\psi_{\alpha\dot\alpha}\nu_{\beta},
\quad
[x_{\alpha\dot\alpha},\bar\theta_{\dot\beta}]_{\star}=-\frac{i}{2}
\psi_{\alpha\dot\alpha}\bar\nu_{\dot\beta},
\\[0.2cm]
[\theta_{\alpha},\theta_{\beta} ]_{\star?}=\frac{i}{4}\varphi_{\alpha\beta}, 
\quad
[\theta_\alpha,\bar\theta_{\dot\beta}]_{\star+}=-\frac{i}{4}\varphi_{\alpha\dot\beta}, 
\quad
[\bar\theta_{\dot\alpha}, \bar\theta_{\dot\beta}]_{\star+}=\frac{i}{4}\bar\varphi_{\dot\alpha\dot\beta},
\end{array}
\end{equation} 
which in turn are directly restored from the invariant P.B's. (\ref{17/1})-(\ref{20/7}). 
The change $ \{ , \}\rightarrow [,]_{\star\mp}$ restores the remaining Moyal brackets originated from the above considered P.B's that together with the brackets (\ref{48/43}) may be used for the studying Lorentz invariant and supersymmetric quantum field models in  non(anti)commutative superspace.

\section{Noncommutativity of the twistor components}

The twistor associated with $\nu_{\alpha}$ and $x_{\alpha\dot\alpha}$  is formed  by the pair  $Z^{\cal A}=(\omega^{\alpha}, \bar\nu_{\dot\alpha})$, where the first twistor element $\omega^{\alpha}$ is  composed  from $\nu_{\alpha}$ and $x_{\alpha\dot\alpha}$
\begin{equation}\label{49/44}
\omega_{\alpha}=ix_{\alpha\dot\alpha}\bar\nu^{\dot\alpha}.
\end{equation} 
The considered Poisson and Moyal  brackets result in  the commutativity between the twistor components  $\omega_{\alpha}$, $\nu_{\beta}$ and their complex conjugate 
\begin{equation}\label{50/44}
\{\omega_{\alpha}, \nu_{\beta}\}=\{\omega_{\alpha}, \bar\nu_{\dot\beta}\}=\{\bar\omega_{\dot\alpha},\nu_{\beta}\}=\{\bar\omega_{\dot\alpha}, \bar\nu_{\dot\beta}\}=0,
\end{equation}
 because of the P.B's. (\ref{17/1}), (\ref{19/3})  and the orthogonality relations 
\begin{equation}\label{51/44}
\varphi_{\alpha\dot\alpha}\nu^{\alpha}=\varphi_{\alpha\dot\alpha}\bar\nu^{\dot\alpha}=0.
\end{equation}
However, $\omega_{\alpha}$ and $\bar\omega_{\dot\alpha}$ have non zero brackets  with 
$x_{\alpha\dot\alpha}$ 
\begin{equation}\label{52/44}
\{x_{\alpha\dot\alpha},\omega_{\beta}\}=\varphi_{\alpha\dot\alpha}\omega_{\beta}-i\bar\eta\psi_{\alpha\dot\alpha}\nu_{\beta},
\quad  
\{x_{\alpha\dot\alpha},\bar\omega_{\dot\beta}\}=\varphi_{\alpha\dot\alpha}\bar\omega_{\dot\beta}-i\eta\psi_{\alpha\dot\alpha}\bar\nu_{\dot\beta}, 
\end{equation}
as well as,  with $\theta_\alpha$ and $\bar\theta_{\dot\alpha}$
 \begin{equation}\label{53/46}
\begin{array}{c}
\{\omega_{\alpha},\theta_\beta \}= -\frac{i}{2}\bar\eta
\varphi_{\alpha\beta}
,\quad
\{\omega_{\alpha},\bar\theta_{\dot\beta} \}= \frac{i}{2}\bar\eta
\varphi_{\alpha\dot\beta}
,\quad
\eta\equiv\theta_{\alpha}\nu^{\alpha},
\end{array}
\end{equation}
 because of the P.B's.  (\ref{18/2})-(\ref{20/7}), (\ref{50/44}). The Grassmannian scalar $\eta$  has zero P.B's. with $\nu,\omega,\theta$
\begin{equation}\label{54}
\{\eta, \nu_{\alpha}\} =\{\eta, \omega_{\alpha} \}=\{\eta,\theta_\alpha\}=0
\end{equation}
and their complex conjugate.
The multiplication of the relations (\ref{52/44}) by $(i\bar\nu^{\dot\alpha})$ together  with using  (\ref{50/44}) and  (\ref{51/44}) yield zero brackets between  the components  $\omega_{\alpha}$ and  $\omega_{\beta}$ of  the same chirality
\begin{equation}\label{55}
\{\omega_{\alpha},\omega_{\beta}\}=
i{\bar\eta}^2\varphi_{\alpha\beta}\equiv0,\quad \{\bar\omega_{\dot\alpha},\bar\omega_{\dot\beta}\}=0,\end{equation}
but yields  zero  brackets with  $\bar\omega_{\dot\beta}$ 
\begin{equation}\label{56}
\{\omega_{\alpha},\bar\omega_{\dot\beta}\}=i\eta\bar\eta\varphi_{\alpha\dot\beta}.
\end{equation}
 Comparing the latter bracket  with  the  bracket (\ref{20/7}) for $\theta_{\alpha}$ and  
$\bar\theta_{\dot\beta}$ we  get the  relation 
\begin{equation}\label{57}
\begin{array}{c}
\{\omega_{\alpha},\bar\omega_{\dot\beta}\}=-4\eta\bar\eta\{\theta_{\alpha},\bar\theta_{\dot\beta}\}
\end{array}
\end{equation}
pointing out  a connection of the $(\omega,\bar\omega)-$noncommutativity with the $(\theta,\bar\theta)$-nonanti-\\commutativity. On the other side, it shows the connection of the twistor  complex structure  with  supersymmetry. Therefore, the choice of $(\theta,\bar\theta)$-nonanticommutative bracket 
induces the $(\omega,\bar\omega)-$noncommutative bracket. Such a correlation of the spin complex structure with supersymmetry and non(anti)commutativity deserves more carefull studying.

\section{ Lorentz invariant brackets in higher dimensions}

The  brackets (\ref{18/2})-(\ref{20/7}) get more compact form in the Majorana representation 
\begin{equation}\label{58}
\nu_{a}={\nu_\alpha\choose \bar\nu ^{\dot\alpha}},
\quad
\theta_{a}={\theta _\alpha\choose \bar\theta^{\dot\alpha}},
\quad
C^{ab}=\left(\begin{array}{cc} \varepsilon^{\alpha\beta}&0\\
0&\bar\varepsilon_{\dot\alpha\dot\beta}
 \end{array}\right),
\quad
\chi^{a}=C^{ab}\chi_{b}                                                                              \end{equation}
for the considered Weyl spinors $\nu_{\alpha},\theta_{\alpha}$, their c.c.  and the charge conjugation matrix $C^{ab}$. Then  the P.B's. (\ref{17/1})-(\ref{20/7}) are  presented in a form suitable for  generalizations. The  P.B's.  (\ref{17/1})-(\ref{19/3}) take the form  
\begin{equation}\label{59}
\{\nu_{a},\nu_{b}\}=0, \quad \{\theta_{a},\nu_{b}\}=0, 
\quad
\{x_{m},\nu_{a}\}=\varphi_{m}\nu_{a},
\end{equation}
where the  real vectors $x_{m}$ and  $\varphi_{m}$ are defined \cite{BW} by the relations 
\begin{equation}\label{60}
\begin{array}{c}
x_{m}=-{1\over 2}(\tilde{\sigma}_{m})^{\dot\alpha\beta}x_{\beta\dot\alpha},
\quad
x_{\alpha\dot\beta}=(\sigma^{m})_{\alpha \dot\beta}{x_m}, 
\\[0.2cm]
\varphi_{m}=-{1\over 2}(\tilde{\sigma}_{m})^{\dot\alpha\beta}\varphi_{\beta\dot\alpha}\equiv
{1\over 4}(\bar\nu\gamma_{m}\nu)
\end{array}
\end{equation}
and  $\gamma_{m}$ are the Dirac matrices in the  Majorana representation.

To rewrite the  rest of the P.B's. in the  Majorana representation it is  convenient to change the  Majorana spinor  $\nu_{a}$  by other Majorana spinor $\lambda_{a}$  
\begin{equation}\label{61}
\lambda_{a}={\lambda_{\alpha}\choose \bar\lambda^{\dot\alpha}}\equiv(\gamma_{5}\nu)_{a},
\quad 
(\gamma_{5})_a{}^b=\left(\begin{array}{cc} -i\delta_\alpha^\beta&0\\
0&i\delta^{\dot\alpha}_{\dot\beta} \end{array}\right)
\end{equation}
 preserving  the form of the P.B's. (\ref{59}). In terms of the real  Majorana spinor $\lambda_{a}$  and  the composed  vectors  $\varphi_{m}$ and  $\psi_{m}$ 
\begin{equation}\label{62}
\varphi_{m}={1\over 4}(\bar\lambda\gamma_{m}\lambda),
\quad 
\psi_{m}=-{1\over 2}(\tilde{\sigma}_{m})^{\dot\alpha\alpha}\psi_{\alpha\dot\alpha}\equiv
-{1\over 2}(\bar\theta\gamma_{m}\lambda)
\end{equation}
the P.B's. (\ref{17/1})-(\ref{20/7}) of the primordial coordinates $x_{m},\theta_{a},\lambda_{a}$ are presented as follow
\begin{equation}\label{63}
\begin{array}{c}
\{\lambda_{a},\lambda_{b}\}=0,
\quad
\{\theta_{a},\lambda_{b}\}=0,
\quad
\{x_{m},\lambda_{a}\}=\varphi_{m}\lambda_{a},
\\[0.2cm] 
\{x_{m},x_{n}\}=-i\psi_{n}\psi_{m},
\quad
\{x_{m},\theta_{a} \}= -\frac{1}{2}\psi_{m}\lambda_{a},
\quad
\{\theta_{a},\theta_{b} \}=-\frac{i}{4}\lambda_{a}\lambda_{b}. 
\end{array}
\end{equation}

The P.B's. of the secondary composite vectors $\psi_{m}$  and  $\varphi_{m}$ (\ref{62}) between themselves  and  with the primordial coordinates  are  presented in the  form
 \begin{equation}\label{64}
\begin{array}{c}
\{x_{m},\psi_{n}\}=\varphi_{m}\psi_{n} + \varphi_{n}\psi_{m},
\quad
\{\psi_{m},\theta_{b} \}= {i\over 2}\varphi_{m}\lambda_{b},
\quad
\{\psi_{m},\lambda_{a} \}=0,
\\[0.2cm]
\{\psi_{m},\psi_{n}\}=-i\varphi_{m}\varphi_{n}, 
\quad
\{\psi_{m},\varphi_{n}\}=0
\end{array}
\end{equation}
and respectively 
\begin{equation}\label{65}
\begin{array}{c}
\{x_{m},\varphi_{n}\}=2\varphi_{m}\varphi_{n},
\quad
\{\theta_{a},\varphi_{m}\}= \{\lambda_{a},\varphi_{m}\}=\{\varphi_{m},\varphi_{n} \}=0.
\end{array}
\end{equation}
The Poisson brackets  (\ref{63})-(\ref{65}) originally derived for $ D=4$ remain to be valid in $D-$dimensional spaces with $D=2,3,4 (mod 8)$, where the Majorana spinors exist. Using the arguments given in the Section 5 one can restore the Moyal brackets  originated  from the P.B's. (\ref{63})-(\ref{65}) in the higher  dimensions by  the simple change $ \{ ,  \}\rightarrow [,]_{\star\mp}$.

\section{Other Lorentz invariant brackets with one spinor}

Using the  Majorana  spinor $\nu_a$ one  can  constuct other  simple supersymmetric and  Lorentz  invariant brackets. One of the  possible invariant  Poisson bracket  might be 
\begin{equation}\label{66/1}
\begin{array}{c} 
\{ F, G \}= F\,[\, -\frac{i}{4}\,(
\stackrel{\leftarrow}{D}\stackrel{\rightarrow}{D}+
\stackrel{\leftarrow}{\bar D}\stackrel{\rightarrow}{\bar D}) + 
c(\stackrel{\leftarrow}{\partial}\stackrel{\rightarrow}
{\Delta} - \stackrel{\leftarrow}{\Delta}\stackrel{\rightarrow}{\partial})\,]\,G
\end{array}
\end{equation}
 which changes  the brackets (\ref{20/7}) to the brackets 
\begin{equation}\label{66/2}
\begin{array}{c}
\{x_{\alpha\dot\alpha},x_{\beta\dot\beta}\}=
i(\varphi_{\alpha\beta}{\bar\theta}_{\dot\alpha}{\bar\theta}_{\dot\beta}+  
{\bar\varphi}_{\dot\alpha\dot\beta}\theta_\alpha\theta_\beta),
\\[0.2cm]
\{x_{\alpha\dot\alpha},\theta_\beta \}= -\frac{1}{2}\varphi_{\alpha\beta}\bar\theta_{\dot\alpha}
\quad
\{x_{\alpha\dot\alpha},\bar\theta_{\dot\beta} \}=-\frac{1}{2}
{\bar\varphi}_{\dot\alpha\dot\beta}\theta_{\alpha},
\\[0.2cm]
\{\theta_{\alpha},\theta_{\beta} \}=\frac{i}{4}\varphi_{\alpha\beta}, 
\quad
\{\theta_\alpha,\bar\theta_{\dot\beta} \}=0, 
\quad
\{\bar\theta_{\dot\alpha}, \bar\theta_{\dot\beta}\}=\frac{i}{4}\bar\varphi_{\dot\alpha\dot\beta}.
\end{array}
\end{equation}
We see that the bracket of the spinor components $\theta_{\alpha}$, ${\bar\theta}_{\dot\alpha}$ having  opposite chiralities is  not  deformed and remains equal zero. Moreover, the brackets $x_{\alpha\dot\alpha}$ with $\theta_{\alpha}$  and  ${\bar\theta}_{\dot\alpha}$ don't preserve the  chiralities of $\theta$  and  $\bar\theta$  spinors.  As a result, one can find breaking of the Jacoby identity for the 
$(x,\theta,\bar\theta) $-cycle, because of  the  relation
\begin{equation}\label{66/2'}
Cycle\{\{ x_{\alpha\dot\alpha}, \theta_{\beta},\bar\theta_{\dot\gamma}\}\}= 
\{\{ x_{\alpha\dot\alpha}, \theta_{\beta},\bar{\theta}_{\dot\gamma}\} + \{\{ x_{\alpha\dot\alpha
}, \bar{\theta}_{\gamma}\},\bar\theta_{\beta}\}
= -\frac{i}{4}\varphi_{\alpha\beta}\bar{\varphi}_{\dot\alpha\dot\beta}.
\end{equation}
So, we conclude that the Lorentz invariant P.B. (\ref{66/1}) has to be excluded. 
But, the next supersymmetric and  Lorentz invariant Poisson bracket
\begin{equation}\label{66/3}
\begin{array}{c} 
\{ F, G \}= F\,[\, \frac{i}{4}\,(
\stackrel{\leftarrow}{D}\stackrel{\rightarrow}{\bar D}+
\stackrel{\leftarrow}{\bar D}\stackrel{\rightarrow}{D}) + 
\frac{1}{2}(\stackrel{\leftarrow}{\partial}\stackrel{\rightarrow}
{\Delta} - \stackrel{\leftarrow}{\Delta}\stackrel{\rightarrow}{\partial})\,]\,G
\end{array}
\end{equation}
is proved to be selfconsistent and yields the following invariant  Poisson brackets
 \begin{equation}\label{66/4}
\begin{array}{c}
\{x_{\alpha\dot\alpha},x_{\beta\dot\beta}\}=
-i(\varphi_{\alpha\dot\beta}{\bar\theta}_{\dot\alpha}\theta_\beta -
\varphi_{\beta\dot\alpha}{\bar\theta}_{\dot\beta}\theta_{\alpha}),
\\[0.2cm]
\{x_{\alpha\dot\alpha},\theta_\beta \}= \frac{1}{2}\varphi_{\beta\dot\alpha}\theta_{\alpha},
\quad
\{x_{\alpha\dot\alpha},\bar\theta_{\dot\beta} \}=\frac{1}{2}
\varphi_{\alpha\dot\beta}\bar\theta_{\dot\alpha},
\\[0.2cm]
\{\theta_{\alpha},\theta_{\beta} \}=\{\bar\theta_{\dot\alpha}, \bar\theta_{\dot\beta}\}=0, 
\quad
\{\theta_\alpha,\bar\theta_{\dot\beta} \}=-\frac{i}{4}\varphi_{\alpha\dot\beta}
 \end{array}
\end{equation}
added by the brackets 
\begin{equation}\label{66/4'}
\begin{array}{c}
\{\nu_{\alpha}, \nu_\beta\}=\{\nu_\alpha,\bar\nu_{\dot\beta}\}=
\{\bar\nu_\alpha, \bar\nu_{\dot\beta}\}=0,
\\[0.2cm]
\{\nu_\alpha,\theta_\beta\}=
\{\nu_\alpha, \bar\theta_{\dot\beta}\}=
\{ \bar\nu_{\dot\alpha},\theta_\beta \}=
\{ \bar\nu_{\dot\alpha},\bar\theta_{\dot\beta}\}=0,
\\[0.2cm]
\{ x_{\alpha\dot\alpha}, \nu_\beta \}= \frac{1}{2}\varphi_{\alpha\dot\alpha}\nu_\beta,
\quad
\{ x_{\alpha\dot\alpha}, \bar\nu_{\dot\beta} \}= \frac{1}{2}\varphi_{\alpha\dot\alpha}\bar\nu_{\dot\beta},
\end{array}
\end{equation}
 One can see that  in contrast to the deformation  (\ref{66/1}) the new deformation (\ref{66/3}) generates zero P.B's. for  the  $\theta_{a}$ components  with the same chiralities.  
The P.B's. (\ref{66/4}),(\ref{66/4'})  satisfy  the Jacobi identities and  deserve to be studied in physical applications. The proof of the Jacobi identities for the P.B's.
(\ref{66/4}) and (\ref{66/4'}) is  analogous  to the proof presented  in the Section 4.

The P.B's. (\ref{66/4}) may be presented in the vector form as follows 
\begin{equation}\label{66/5'}
\begin{array}{c}
\{x_{m},x_{n}\}=-\frac{i}{4}(\chi_{m}\bar\chi_{n}-\chi_{n}\bar\chi_{m}),
\\[0.2cm]
\{x_{m},\theta_{\beta}\}=-\frac{1}{4}{\bar\chi}_{m}\nu_{\beta}, \,\, \,
\{x_{m},{\bar\theta}_{\dot\beta}\}=-\frac{1}{4}\chi_{m}{\bar\nu}_{\dot\beta},
\\[0.2cm]
\{\theta_{a},\theta_{b}\}=-\frac{i}{8}(\nu^{(+)}_{a}\nu^{(-)}_{b}+\nu^{(+)}_{b}\nu^{(-)}_{a}),
\end{array}
\end{equation}
where we introduced  the complex Grasssmannian vector $\chi_{m}$  with the real and imaginary parts  presented by $\psi_{1m}, \psi_{2m}$  and the chiral components  $\theta^{(\pm)}$ and $\nu^{(\pm)}$ 
\begin{equation}\label{66/5''}
\begin{array}{c}
\chi_{m}\equiv
(\nu\sigma_{m}\bar\theta)\equiv-\bar\nu\gamma_{m}\frac{1+i\gamma_{5}}{2}\theta\equiv \psi_{1m}+ i\psi_{2m},
\\[0.2cm]
\bar\chi_{m}\equiv(\chi_{m})^{*}= -\bar\nu\gamma_{m}\frac{1-i\gamma_{5}}{2}\theta,
\quad
\psi_{1m}\equiv -\frac{1}{2}(\bar\theta\gamma_{m}\nu),
\quad
\psi_{2m}\equiv -\frac{1}{2}(\bar\theta\gamma_{m}\gamma_{5}\nu),
\\[0.2cm]
\theta^{(\pm)}\equiv\frac{1}{2}
(1 \pm i\gamma_{5})\theta, \,\,\,\,
\nu^{(\pm)}\equiv\frac{1}{2}(1 \pm  i\gamma_{5})\nu .
\end{array}
\end{equation}

Then the  P.B's. (\ref{66/5'}) are presented in the form directly generalizing the P.B's.  (\ref{63})
\begin{equation}\label{66/5'''}
\begin{array}{c}
\{x_{m},x_{n}\}=-\frac{i}{2}(\psi_{1m}\psi_{1n} + \psi_{2m}\psi_{2n}),
\\[0.2cm]
\{x_{m},\theta_{a}\}=-\frac{1}{4}(\psi_{1m}\nu_{a} + \psi_{2m}\lambda_{a}),
\\[0.2cm]
\{\theta_{a},\theta_{b}\}=-\frac{i}{8}(\nu_{a}\nu_{b}+\lambda_{a}\lambda_{b}),
\end{array}
\end{equation}
where $\lambda_{a}\equiv(\gamma_{5}\nu)_{a}$ (\ref{61}).
Comparing (\ref{66/5'''}) with (\ref{63}) we observe that the change of the P.B. (\ref{13/31}) by 
(\ref{66/3}) is equivalent to the complexification of the real Grassmannian vector $\psi_{m}$ (\ref{62}) accompanied by the appearance of the spinors $\nu_{a}$ and $(\gamma_{5}\nu)_{a})$ in the r.h.s. of  (\ref{66/5'''}).

The P.B's. (\ref{66/3}) may be generalized to the case of extended supersymmetries with $N>1$ 
\begin{equation}\label{66/5}
\begin{array}{c} 
\{ F, G \}= F\,[\, \frac{i}{4}\,(
\stackrel{\leftarrow}{D_{i}}\stackrel{\rightarrow}{\bar D^{i}}+
\stackrel{\leftarrow}{\bar D^{i}}\stackrel{\rightarrow}{D_{i}}) + 
\frac{1}{2}(\stackrel{\leftarrow}{\partial}\stackrel{\rightarrow}
{\Delta} - \stackrel{\leftarrow}{\Delta}\stackrel{\rightarrow}{\partial})\,]\,G,
\end{array}
\end{equation}
where $D_{i}=\nu_\alpha D^{\alpha}_{i}$ and ${\bar D}^{i}={\bar\nu}_{\dot\alpha}
{\bar D}^{\dot\alpha i}\,,  (i=1,2,..,N)$. 
The P.B's.  (\ref{66/5}) generate  the following brackets  for the primordial space-time  (super)coordinates 
\begin{equation}\label{66/6}
\begin{array}{c}
\{x_{\alpha\dot\alpha},x_{\beta\dot\beta}\}=
-i(\varphi_{\alpha\dot\beta}{\bar\theta}_{\dot\alpha i}\theta^{i}_{\beta} -
\varphi_{\dot\alpha\beta}{\bar\theta}_{\dot\beta i}\theta^{i}_{\alpha}),
\\[0.2cm]
\{x_{\alpha\dot\alpha},\theta^{i}_\beta \}= \frac{1}{2}\varphi_{\dot\alpha\beta}\theta^{i}_{\alpha},
\quad
\{x_{\alpha\dot\alpha},\bar\theta_{\dot\beta i} \}=\frac{1}{2}
\varphi_{\alpha\dot\beta}\bar\theta_{\dot\alpha i},
\\[0.2cm]
\{\theta^{i}_{\alpha},\theta^{k}_{\beta} \}=\{\bar\theta_{\dot\alpha i}, \bar\theta_{\dot\beta k}\}=0, 
\quad
\{\theta^{i}_{\alpha},{\bar\theta}_{\dot\beta k } \}=\frac{i}{4}\varphi_{\alpha\dot\beta}.
{\delta^{i}}_{k}
\end{array}
\end{equation}
The rest of the P.B's for $x_{\alpha\dot\alpha},\nu_{a},\theta^{i}_{\alpha}$ coincides with the  P.B's. (\ref{66/4'}).

\section{More spinors - more Lorentz invariant brackets}

The above consideration  to construct  the  Lorentz covariant Poisson and Moyal brackets was  restricted by the simplest case of one  additional spinor coordinate which resulted in the appearance of the supersymmetric derivatives $ D^{\alpha},{\bar D}^{\dot\alpha}$ and $\partial^{\alpha\dot\alpha}$ (\ref{3/30}) in the considered P.B's. only in the form of the scalars (\ref{5/32}). Using only these scalars for the construction of  the invariant  P.B. (\ref{13/31}) restricts  the class of admissible motions in superspace. To extend this class still preserving 
 the Lorentz invariance and supersymmetry one can introduce additional independent spinor coordinates. In the case $D=4$ it is enough to add only  one new spinor coordinate $\mu_{\alpha}$, because $\mu_{\alpha}$ and  $\nu_{\alpha}$ form the complete  spinorial basis  and may be identified with the Newman-Penrose dyad \cite{Penrose}
\begin{equation}\label{66}
\mu^\alpha \nu_\alpha\equiv \mu^\alpha\varepsilon_{\alpha\beta}\nu^\beta=1,
\quad
\mu_\alpha\nu_\beta - \mu_\beta\nu_\alpha=\varepsilon_{\alpha\beta}.
\end{equation}
Then one  can  form four independent Lorentz  invariant  supersymmetric differential operators 
\begin{equation}\label{67}
\begin{array}{c}
D^{(\nu)}= \nu_{\alpha}D^{\alpha},
\quad 
\bar D^{(\nu)}= \bar\nu_{\dot\alpha}\bar{D^{\dot\alpha}},
\quad 
D^{(\mu)}= \nu_{\alpha}D^{\alpha},
\quad 
\bar D^{(\mu)}= \bar\mu_{\dot\alpha}\bar{D^{\dot\alpha}},
\end{array}
\end{equation}
two of which $D^{(\nu)},\bar D^{(\nu)}$ coincide with the operators $D,\bar D$ (\ref{5/32}).
 Their linear combinations 
\begin{equation}\label{68}
\begin{array}{c} 
D^{(\nu)}_{\pm}\equiv D^{(\nu)}\pm \bar D^{(\nu)},
\quad 
D^{(\mu)}_{\pm}\equiv D^{(\mu)}\pm \bar D^{(\mu)},
\end{array}
\end{equation}
 form four Lorentz  invariant and supersymmetric  supersubalgebras
\begin{equation}\label{69}
\begin{array}{c} 
[D^{(\nu)}_{\pm}, D^{(\nu)}_{\pm}]_{+}=
\mp8i\partial^{(\nu)},
\quad 
[ D^{(\nu)}_{\pm},  \partial^{(\nu)})]=
[ \partial^{(\nu)},\partial^{(\nu)}]=0,
\quad
\partial^{(\nu)}\equiv(\nu_{\alpha}{\bar\nu}_{\dot\alpha} \partial^{\alpha\dot\alpha}),
\\[0.2cm] 
[D^{(\mu)}_{\pm}, D^{(\mu)}_{\pm}]_{+}=
\mp8i\partial^{(\nu)},
\quad 
[ D^{(\mu)}_{\pm},  \partial^{(\mu)})]=
[ \partial^{(\mu)},\partial^{(\mu)}]=0,
\quad
\partial^{(\mu)}\equiv(\mu_{\alpha}{\bar\mu}_{\dot\alpha} \partial^{\alpha\dot\alpha}),
\end{array}
\end{equation}
which are connected by the P.B. relations 
\begin{equation}\label{70}
\begin{array}{c}
[D^{(\nu)}_{\pm}, D^{(\mu)}_{\pm}]_{+}=\mp4i \partial^{(+)},
\quad 
\partial^{(+)}\equiv(\nu_{\alpha}{\bar\mu}_{\dot\alpha}+ \mu_{\alpha}{\bar\nu}_{\dot\alpha})
 \partial^{\alpha\dot\alpha},
\\[0.2cm] 
[D^{(\nu)}_{\pm}, D^{(\mu)}_{\mp}]_{+}=\pm4i \partial^{(-)},
\quad 
\partial^{(-)}\equiv(\nu_{\alpha}{\bar\mu}_{\dot\alpha}- \mu_{\alpha}{\bar\nu}_{\dot\alpha})
 \partial^{\alpha\dot\alpha}.
\end{array}
\end{equation}
It is  easy  to see that the Lorentz invariant and supersymmetric differential operators  $D^{(\nu)}_{\pm},  D^{(\mu)}_{\pm},\partial^{(\nu)},\partial^{(\mu)},\partial^{(\mp)}$ describe whole class of  admissible   motions in the superspace and  together with the extended dilatation operator  $\Delta'$
\begin{equation}\label{71}
\begin{array}{c}
\Delta'=(\nu_{\alpha}
\frac{\partial}{\partial\nu_\alpha} + \bar\nu_{\dot\alpha}\frac {\partial}{\partial\bar\nu_{\dot\alpha}})
-
(\mu_{\alpha}
\frac{\partial}{\partial\mu_\alpha} + \bar\mu_{\dot\alpha}\frac {\partial}{\partial\bar\mu_{\dot\alpha}})
\end{array}
\end{equation}
preserving the condition (\ref{66}) may be  used as invariant bilding blocks for the construction of more general Lorentz invariant supersymmetric Poisson and Moyal brackets. 

Then the  Lorentz invariant  and supersymmetric Poisson  bracket 
\begin{equation}\label{72}
\begin{array}{c} 
\{ F, G \}= F\,[\, -\frac{i}{4}\,(
{\stackrel{\leftarrow}{D}}^{(\nu)}_{-}
{\stackrel{\rightarrow}{D}}^{(\nu)}_{-}+ 
{\stackrel{\leftarrow}{D}}^{(\mu)}_{-}
{\stackrel{\rightarrow}{D}}^{(\mu)}_{-})+
c(\stackrel{\leftarrow}{\partial^{(\nu)}}+\stackrel{\leftarrow}{\partial^{(\mu)}})
\stackrel{\rightarrow}{\Delta'} - 
\stackrel{\leftarrow}{\Delta'}(\stackrel{\rightarrow}{\partial^{(\nu)}}
+\stackrel{\rightarrow}{\partial^{(\mu)}})
\,]\,G .
\end{array}
\end{equation}
 might  be considered as a candidate for the  generalizations of (\ref{13/31}). The P.B. (\ref{72})
yields  the following  coordinate  P.B's.
\begin{equation}\label{73}
\begin{array}{c}
\{x_{m},x_{n}\}=-i(\psi^{(\nu)}_{n}\psi^{(\nu)}_{m} +\psi^{(\mu)}_{n}\psi^{(\mu)}_{m}),
\\[0.2cm] 
\{x_{m},\theta_{a} \}= -\frac{1}{2}(\psi^{(\nu)}_{m}\lambda^{(\nu)}_{a}+ 
\psi^{(\mu)}_{m}\lambda^{(\mu)}_{a}),
\\[0.2cm]
\{\theta_{a},\theta_{b} \}=-\frac{i}{4}(\lambda^{(\nu)}_{a}\lambda^{(\nu)}_{b} + 
\lambda^{(\mu)}_{a}\lambda^{(\mu)}_{b}) 
\end{array}
\end{equation}
for the primordial coordinates $x_m$ and $\theta_a$, where the additional Majorana spinor $\lambda^{(\mu)}_{a}$ and the Grassmannian vector $\psi^{(\mu)}_{n}$ are defined by the relations
\begin{equation}\label{74}
\begin{array}{c}
\quad
\psi^{(\nu)}_{n} \equiv\psi_{n},
\quad 
\lambda^{(\nu)}_{a}\equiv\lambda_{a},
\quad 
\psi^{(\mu)}_{n}\equiv
{1\over 2}(\bar\theta\gamma_{n}\lambda^{(\mu)}),
\quad 
\lambda^{(\mu)}_{a}\equiv(\gamma_{5}\mu)_{a}
\end{array}
\end{equation}
The primordial Majorana spinors $\lambda^{(\nu)}_{a}$ and $\lambda^{(\mu)}_{a}$ have zero P.B's. between themselves and with $\theta_a,\psi^{(\nu)}_{m}, \psi^{(\mu)}_{n}$,  but non zero  P.B's. with  $x_{m}$ 
\begin{equation}\label{75}
\begin{array}{c}
\{x_{m},\lambda^{(\nu)}_{a}\}=c(\varphi^{(\nu)}_{m}+\varphi^{(\mu)}_{m})\lambda^{(\nu)}_{a}),
\quad
\{x_{m},\lambda^{(\mu)}_{a}\}=-c(\varphi^{(\nu)}_{m}+\varphi^{(\mu)}_{m})\lambda^{(\mu)}_{a},
\\[0.2cm] 
\varphi^{(\nu)}_{m}\equiv\varphi_{m},\quad
\varphi^{(\mu)}_{m}\equiv{1\over 4}(\bar\lambda^{(\mu)}\gamma_{m}\lambda^{(\mu)}),
\end{array}
\end{equation}
where the real constant  $c$ has to be defined from the solution of the Jacobi identities. 
We intend to give  back to the  studying this P.B.  and  other possible  generalizations of  the P.B's (\ref{63}) in other place. 

\section{Conclusion}

It was shown here that the extension of the  $N=1$ superspace ($x_m, \theta_a$) by commuting Majorana spinors may be used for the construction of supersymmetric and Lorentz invariant Poisson and Moyal brackets generating deformed non(anti)commutative relations for space-time (super)coordinates.
To make clear the proposal we elaborate the case of one additional spinor $\lambda_{a}$ extending the standard  $N=1$ superspace to the non(anti)commutative superspaces free of background  fields.
The corresponding Lorentz invariant and supersymmetric coordinate brackets were presented.
It was established that noncommutativity of $x_m$ with  $x_n$ and $\theta_a$ is measured  by the real or complex  Grassmannian vectors $\psi_{m}$ composed from $\theta_a$ and $\lambda_{a}$, 
which are  known as  dynamical variables describing the spin degrees of freedom of spinning string 
or  particle. At the same time, the  nonanticommutativity of  the $\theta_a$ componets between themselves is measured by only additional spinor or its chiral components. These results hint on a hidden spinorial structure of space-time encoded in the Penrose twistor picture and its supersymmetric extensions as an alternative source of (super)coordinate non(anti)commutativity. In the simplest case corresponding to the P.B's. (\ref{63}) a  correlation between the spinorial structure and supergravity fields may be schematically illustrated by the  correspondence:
\begin{equation}\label{76}
\begin{array}{c}
\theta_{mn}\leftrightarrow i\psi_{m}\psi_{n},\quad
C_{ab}\leftrightarrow \lambda_{a}\lambda_{b},\quad
 \Psi^{a}_{m}\leftrightarrow \psi_{m}\lambda^{a},
 \end{array}
\end{equation}
where  $B_{mn}=\theta^{-1}_{mn}$, $C_{ab}$ and $\Psi^{a}_{m}$ are constant antisymmetric field, the graviphoton and the gravitino respectively. The map (\ref{76}) transforms the well-known field dependent bracket relations into the brackets  (\ref{63}) and vice versa.
On the other hand, such a correspondence hints on a connection with the known Feynman-Wheeler picture and its superymmeric generalization \cite{TZ}, where the  Maxwell supermultiplet fields arise as secondary objects constructed from the superspace coordinates.
We outlined  also a  way  to  construction of more general supersymmetric Lorentz invariant brackets  for  the  cases of $N$ extended  supersymmetries  and  additional spinor coordinates  using Lorentz  invariant  supersymmetric derivatives generalizing  (\ref{5/32}). Studying that generalizations and the corresponding  deformations of quantum field models are  under consideration. 

\section{Acknowledgements}

The author  thanks Fysikum at the Stockholm University for kind hospitality and I. Bengtsson and   B. Sundborg for useful discussions. The work was partially supported by the grant of the Royal Swedish Academy of Sciences and by the SFFR of Ukraine under Project 02.07/276.

\end{document}